\documentstyle[12pt]{article}
\newlength{\dinwidth}
\newlength{\dinmargin}
 \newcommand\noi{\noindent}
 \newcommand\beq{\begin{equation}}
 \newcommand\eeq{\end{equation}}
 \newcommand\beqn{\begin{eqnarray}}
 \newcommand\eeqn{\end{eqnarray}}
 \newcommand{\doublespace} {
 \renewcommand{\baselinestretch} {1.6}
 \large\normalsize}
\newcommand{\la}{\langle}
 \newcommand{\ra}{\rangle} 

\begin{document}
\vspace*{1cm}
\hspace*{9cm}{\Large MPI H-V28-1996}
\vspace*{3cm}

\begin{center}
{
\renewcommand{\thefootnote}{\fnsymbol{footnote}}

{\Large \bf Interplay of Soft and Hard Interactions}\\
%
{\Large {\bf in Nuclear Shadowing at 
High $Q^2$ and Low $x$}\footnote{
To appear in the Proceedings of the Workshop on Future Physics at HERA,\\
DESY, September 25, 1995 -- May 31, 1996}}
}

\vspace*{5mm}

{\large Boris~Kopeliovich$^{ab}$
and Bogdan Povh$^a$}\\
\end{center}
$^a$ Max-Planck Institut f\"ur
 Kernphysik, Postfach
103980,  69029 Heidelberg, Germany\\
$^b$ Joint Institute
 for Nuclear Research,
Dubna, 141980
 Moscow Region, Russia\\
\vspace*{1cm}
{{
\doublespace
\begin{quotation}
\noindent
{\bf Abstract:}
Nuclear shadowing corrections are dominated by soft
interaction and grow as function of $1/x$ more slowly than
the single scattering term, which has an essential
contribution from hard interaction. Therefore
we predict vanishing nuclear shadowing at very low
$x$ provided that $Q^2$ is high and fixed. At the same time,
at medium and low $Q^2$, nuclear shadowing grows
with $1/x$ as is well known for soft hadronic interactions.
\end{quotation}
\newpage

Experimental observation \cite{nmc1,nmc2} of nuclear shadowing 
in deep-inelastic scattering at low $x$ was probably the first signal
that this process is substantially contaminated by
soft physics even at high $Q^2$. Since nuclear shadowing 
is closely related to diffraction \cite{gribov}, it is not surprising
that recent measurements at HERA found diffraction to be 
a large fraction of the total cross section.

The structure function $F_2(x,Q^2)$ is proportional to the
total cross section of interaction of a virtual photon with the target.
This invites one to consider deep-inelastic lepton scattering
in the rest frame of the target, where the virtual photon 
demonstrates its hadronic properties. Namely, the hadronic fluctuations 
of the photon interact strongly with the target \cite{bauer}. 
Such a process looks quite different from the 
partonic interpretation of deep-inelastic scattering.
The observables are Lorentz-invariant,
but the space-time interpretation depends on
the reference frame.

The observed virtual photoabsorption cross 
section on a nucleon is an
average of total interaction cross sections 
$\sigma^{hN}_{tot}$ of  
hadronic fluctuations  weighted by probabilities 
$W^{\gamma^*}_h$,
\beq
\sigma^{\gamma^*N}_{tot}(x,Q^2) = 
\sum\limits_h W^{\gamma^*}_h(x,Q^2)\ 
\sigma^{hN}_{tot} \equiv 
\la \sigma^{hN}_{tot}\ra\ .
\label{1}
\eeq
In the case of a nuclear target 
the same procedure leads to \cite{kp},
\beq
\frac{\sigma^{\gamma^*A}_{tot}(x,Q^2)}
{\sigma^{\gamma^*N}_{tot}(x,Q^2)} = 
1-{1\over 4}\frac{\la(\sigma^{hN}_{tot})^2\ra}
{\la \sigma^{hN}_{tot}\ra}\ 
\la T\ra\ F_A^2(q_L) + ...
\label{2}
\eeq
\noi
Here $T(b) = \int_{-\infty}^{\infty}dz\ \rho_A(b,z)$ is 
the nuclear thickness at impact parameter $b$ and 
$\la T\ra = (1/A)\ \int d^2b\ T^2(b)$ is its mean value.
$\rho_A(b,z)$ is the nuclear density dependent on $b$ and
longitudinal coordinate $z$.
The longitudinal nuclear formfactor 
\beq
F^2_A(q_L) = 
\frac{1}{A\la T\ra}\ 
\int d^2b\ \left|\int\limits_{-\infty}^{\infty}
dz\ \rho_A(b,z)\ \exp(iq_L z)\right|^2
\label{3}
\eeq
\noi
takes into account the 
effects of the finite lifetime $t_c\approx 1/q_L$ 
(the coherence time) of 
hadronic fluctuations of the photon, where,
$q_L = (m_h^2 + Q^2)/2m_N\nu$ is the 
longitudinal momentum transfer in 
$\gamma^* N \to h N$. At large $q_L > 1/R_A$, the nuclear 
formfactor (\ref{3}) vanishes and suppresses the shadowing
term (\ref{2}). This is easily interpreted: for large $q_L$
the fluctuation lifetime and its path 
in nuclear medium are shorter, and shadowing is reduced.
For further estimations we assume that the mean mass squared
of a photon fluctuation is $Q^2$, leading
to $q_L = 2m_Nx$.

Thus, all the factors in the first-order shadowing term
(\ref{2}) are known, except $\la(\sigma^{hN}_{tot})^2\ra$.
First, let us analyse the $Q^2$-behaviour of this factor. 
It is known that $\la\sigma^{hN}_{tot}\ra \propto 1/Q^2$ 
according to Bjorken scaling.
In QCD this is usually interpreted as a consequence of color screening:
the higher the value of $Q^2$, the smaller the mean transverse
size squared $\la\rho^2\ra \sim 1/Q^2$ of its hadronic fluctuation.
Due to color screening the cross section of interaction of 
such a fluctuation with external gluonic fields vanishes as 
$\sim 1/Q^2$. However, the situation is more
complicated, as a finite admixture of soft fluctuations having 
large size is unavoidable \cite{bk,fs,nz91}. 
We classify in a simplified way the hard and soft mechanisms
of deep-inelastic scattering in Table~1.

\vspace{.2cm}
\begin{center}
Table 1.
Contributions of soft and  hard fluctuations
of a virtual photon\\ 
to the DIS cross section
and to nuclear shadowing

\vspace{.3cm}
{\doublespace

\begin{tabular}{|c|l|l|l|l|c|}
 \hline 
Fluctuation & $W^{\gamma*}_h$ & $\sigma^{hN}_{tot}$ & 
$W^{\gamma*}_h\sigma^{hN}_{tot}$ &
 $W^{\gamma*}_h(\sigma^{hN}_{tot})^2$\\
 \hline Hard & $\sim 1$ & $\sim 1/Q^2$ 
& $\sim 1/Q^2$  & $\sim 1/Q^4$\\
 \hline
 Soft & $\sim\mu^2/Q^2$ &$\sim1/\mu^2$& 
$\sim 1/Q^2$ & $\sim 1/\mu^2Q^2$\\
\hline
\end{tabular}
}
\vspace{.3cm}
\end{center}

As previously stated, the mean fluctuation of a highly virtual
photon is hard and has a small transverse size
$\sim 1/Q^2$. This is why we assign to it  
a weight close to $1$ and a small $\sim 1/Q^2$ 
cross section in Table~1. 
On the contrary, a soft fluctuation having a large 
size $\sim 1/\mu^2$, where $\mu$ is a soft parameter 
of the order of $\Lambda_{QCD}$, is expected to be
quite rare in the photon, suppressed by factor 
$\sim \mu^2/Q^2$.\footnote[1]{This is shown to be true 
for a transverse
photon by perturbative calculations \cite{nz91}, but
in a longitudinal photon soft fluctuations have an
extra suppression $\sim 1/Q^4$ \cite{nz91}. Therefore
shadowing is a higher twist effect.} On the other hand, 
such a soft fluctuation has a large $\sim 1/\mu^2$
cross section. Therefore, the soft contribution
to  $\la\sigma^{hN}_{tot}\ra$ has the same leading
twist behaviour $\sim 1/Q^2$ as the hard one. 
Thus, according to Table~1 one cannot say that
Bjorken scaling results only from the smallness of the interaction
cross section of hard
photon fluctuations, that also arises from the rareness 
of the soft components of a virtual photon.

The last column of Table~1 summarizes the $Q^2$-dependence of
hard and soft contributions to $(\la\sigma^{hN}_{tot})^2\ra$,
which sets the size of nuclear shadowing and diffraction effects. 
In this case the hard component turns out to be a higher twist effect, 
and the leading contribution comes from soft interaction\footnote[2]{
Soft interaction also contributes to the higher twist terms 
\cite{prw}, which we neglect, provided that $Q^2$ is sufficiently high.}.
This is why the applicability of pure perturbative calculations 
to nuclear shadowing or diffraction is questionable.

This conclusion is different from the statement in paper \cite{glm}
that quite a small transverse size, $\sim 0.2\ fm$, is typical for 
a hadronic fluctuation
of a transversely polarized photon in diffractive dissociation.
This value corresponds to a tiny cross section $\sim 1\div 1.5\ mb$,
which is in a strong disagreement 
with the observed nuclear shadowing \cite{nmc1,nmc2}, 
which demands $\la(\sigma_{tot}^{hN})^2\ra/
\la\sigma_{tot}^{hN}\ra \sim 20\ mb$.

Since $(\la\sigma^{hN}_{tot})^2\ra$ is
dominated by soft interactions, we can parameterize 
it as \cite{kp},
\beq
\frac{(\la\sigma^{hN}_{tot})^2\ra}
{\la\sigma^{hN}_{tot}\ra} = 
\frac{N}{F_2^p(x,Q^2)}\ 
\left({1\over x}\right)^{2\Delta_P(\mu^2)}
\label{5}
\eeq
\noi
We are interested in the behaviour
of this factor at very low $x$ and assume
for the numerator dominance of
the soft Pomeron with intercept
$\alpha_P(0) = 1 + \Delta(\mu^2)$, where
$\Delta(\mu^2) \approx 0.1$ is known from Regge
phenomenology of soft hadronic interactions.
This explains particularly why a soft 
Pomeron intercept was found 
in diffraction at high $Q^2$ \cite{h1}, 
while the intercept describing $x$-dependence of $F_2^p(x,Q^2)$
at high $Q^2$ is much larger, $\Delta_P(Q^2) \approx 
0.3 \div 0.4$ \cite{kp}.

The proton structure function was fixed in \cite{kp} by
fitting available data. The only unknown parameter $N$
is universal for all nuclei and is fixed by the fit at
$N=3\ GeV^{-2}$ \cite{kp}.

Now we are in a position to predict nuclear shadowing down to low $x$.
The fact that $2\Delta_P(\mu^2) < \Delta_P(Q^2)$ at high $Q^2$
leads to the unusual prediction of vanishing nuclear 
shadowing at very low $x$. That is, the first shadowing correction
in (\ref{2}) decreases with $1/x$ provided that the nuclear 
formfactor saturates, $F_A^2(x) \to 1$. This is demonstrated in
Figs.~1-2, where we have plotted our predictions for carbon and tin
versus $x$ at fixed values of $Q^2$. 
Note that formula (\ref{2}) does not include
the small (a few percent) effect of nuclear antishadowing.
We have renormalized all curves
by factor 1.03 in order to incorporate this effect, which we
assume to be A-independent for simplicity. 
}}
\begin{figure}[tbh]
\includegraphics{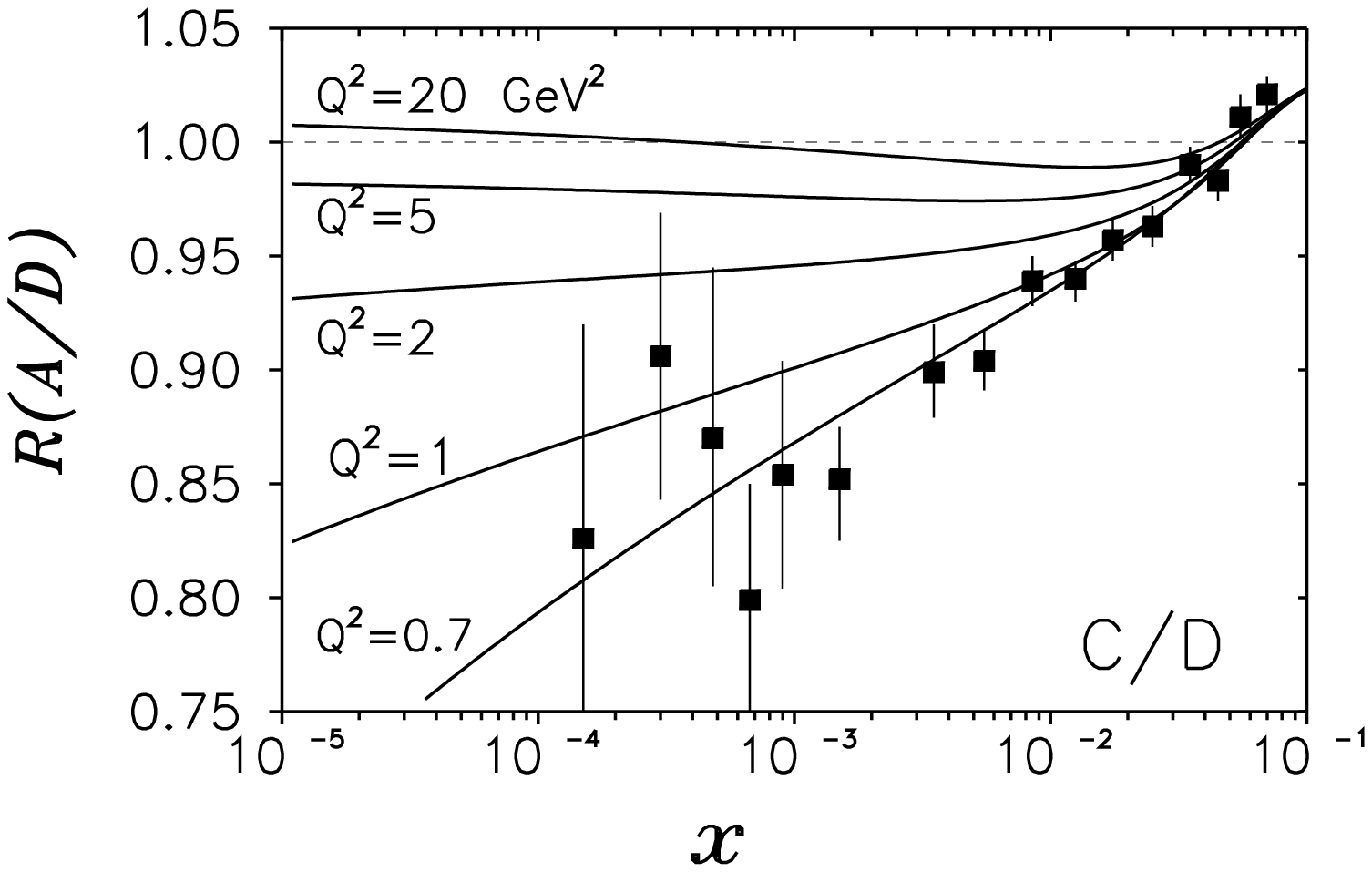}
\includegraphics{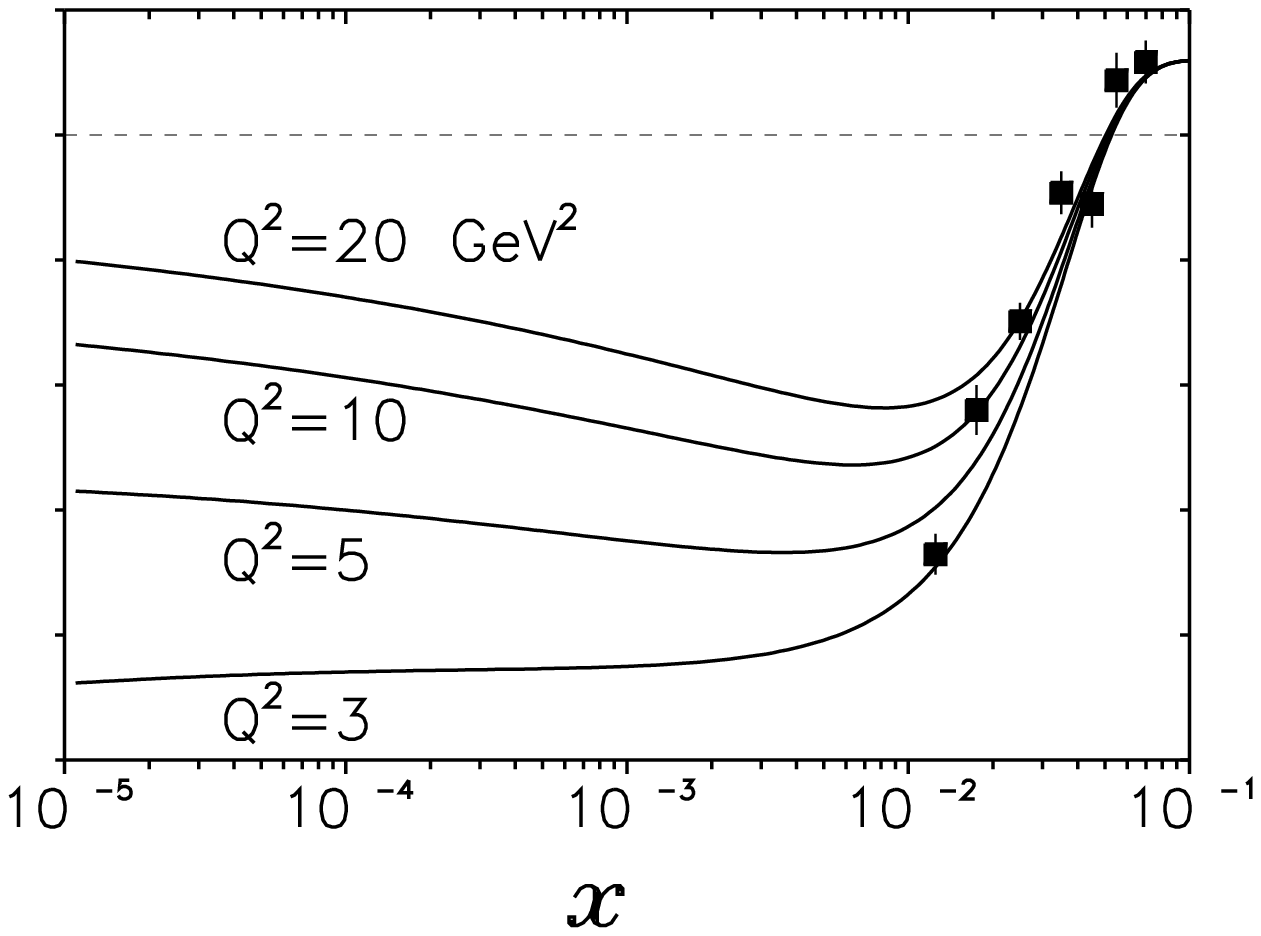}
\begin{center}
\vspace{7cm}
\parbox{13cm}
{\caption[Delta]
{\it Normalized ratio $F_2^A(x,Q^2)/F_2^D(x,Q^2)$ 
calculated for carbon using (\ref{2}).
The data points are from 
\cite{nmc1}.}
\label{fig1}}
{\caption[Delta]
{\it The same as in Fig.~1 but for tin. The data points are from 
\cite{nmc3}.}
\label{fig2}}
\end{center}
\end{figure}
{{\doublespace
Note that comparison with 
data \cite{nmc1} in Fig.~1 is marginal, since $Q^2$ substantially 
varies from point to point. To make the comparison 
with data more sensible it was suggested
in \cite{kp} to plot the data against a new variable,
\beq
 n(x,Q^2,A)={1\over 4}
 \frac{N}{F_2^p(x,Q^2)}
 \langle
 T(b)\rangle F^2_A(q_L) \left({1\over
 x}\right)^{2\Delta_P(\mu^2)}\ .
\label{6}
\eeq
One may expect according to (\ref{2}) - (\ref{5}) that
nuclear shadowing is $A$-, $x$- and $Q^2$-independent at
fixed $n(x,Q^2,A)$.
Data from the NMC experiment plotted against $n(x,Q^2,A)$
in Fig.~3 nicely confirm such a scaling.
Note that the data points for different nuclei may
differ within a few percent due to the antishadowing effect,
if it is A-dependent.
}}
\begin{figure}[tbh]
\includegraphics{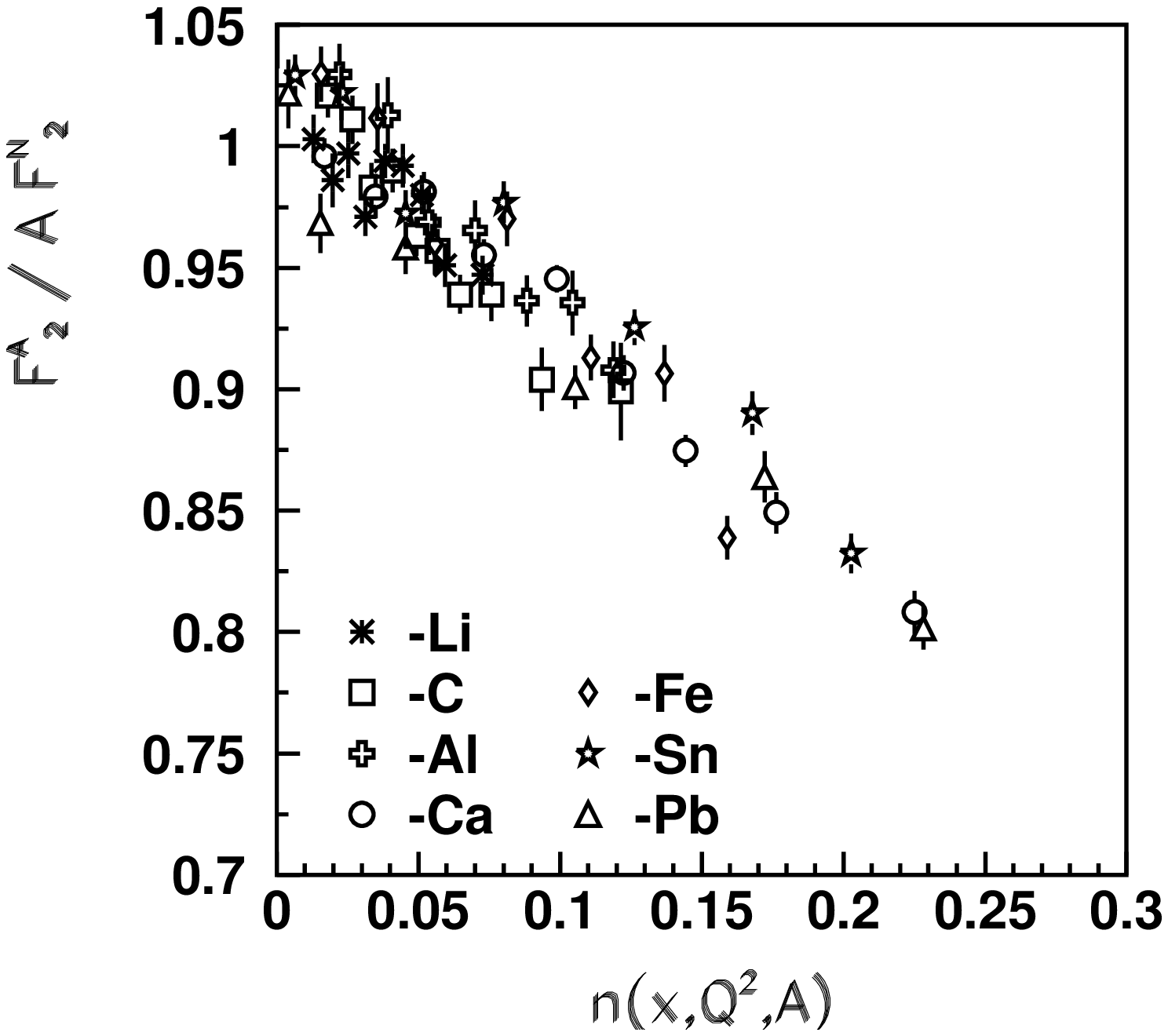}
\includegraphics{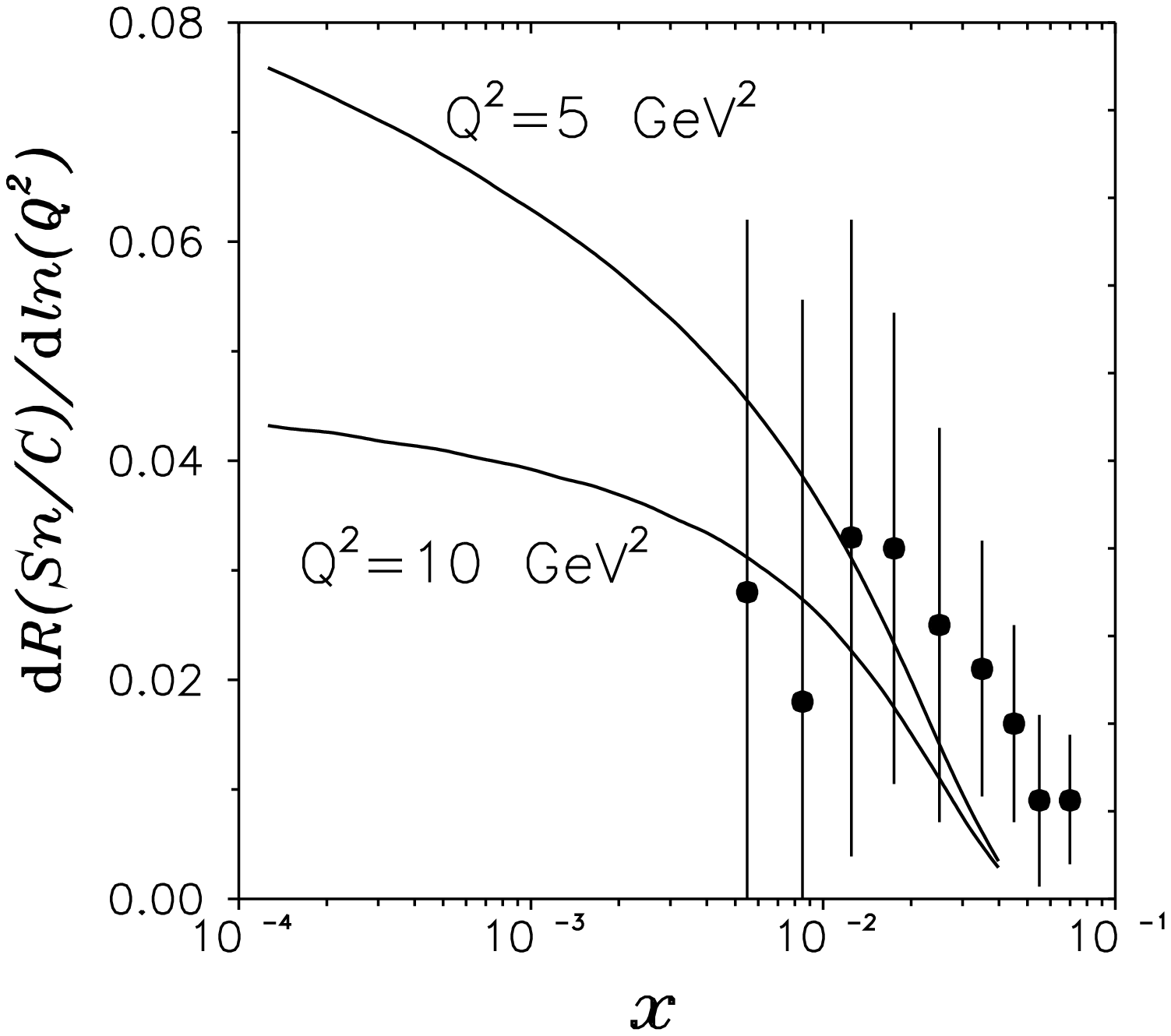}
\begin{center}
\vspace{7cm}
\parbox{13cm}
{\caption[Delta]
{\it Nuclear shadowing versus scaling variable $n(x,Q^2,A)$ (see text).
The data for Li, C and Ca are from \cite{nmc1,nmc2}, other data
points are from \cite{nmc3}.}
\label{fig3}}
\parbox{13cm}
{\caption[Delta]
{\it Logarithmic $Q^2$-derivative of the ratio of structure
functions for tin to carbon. The data are from \cite{nmc4}.}}
\end{center}
\end{figure}
{{\doublespace
The results depicted in Figs.~1-2 demonstrate a
substantial 
variation of nuclear shadowing 
with $Q^2$, especially at low $x$. $Q^2$-dependence of shadowing
was observed recently by the NMC experiment \cite{nmc4}.
Their data are plotted in Fig.~3 together with our predictions,
which reproduce well the order of magnitude of the effect.
We cannot claim a precise description, since the data
represent the results of averaging over large interval of $Q^2$ down
to quite low values, where our approximation may not
work. It is important that all $Q^2$-dependence of
shadowing originates in formula (\ref{2}) only from the
proton structure function in the denominator. Consequently, this
effect in (\ref{2}) has no relation to shadowing of gluons in
nuclei.

To conclude, we would like to comment on the approximations used.

First of all, in saying that $\la\sigma^{hN}_{tot}\ra$
is dominated by soft interaction we neglected the higher
twist corrections $\sim 1/Q^2$ presented in Table~1. Thus,
one should be cautious using this approximation at low $Q^2$.

Although we use a double-leading-log type parameterization
for $F_2^p(x,Q^2)$, which provides a vanishing effective 
$\Delta_P$ at $x \to 0$, it is almost a constant in the
range of $x$ under discussion, i.e. is compatible with
the BFKL solution \cite{bfkl}.

It is easy to show that the higher-order shadowing corrections omitted in
(\ref{2}) are soft as well. However, the $x$-dependence of
the $n$-fold correction is governed by the power $n\Delta_P(\mu^2) -
\Delta_P(Q^2)$ which may be positive for $n=3$ or $4$ and so on.
A question arises, whether the growth of higher-order shadowing
corrections can change our conclusion about the shadowing 
decreasing with 
$1/x$. We think it cannot. Indeed, let us consider an
eikonal shadowing where the first term correspond to the hard Pomeron
with a large $\Delta_P(Q^2)$, but all other terms correspond
to the soft $\Delta_P(\mu^2)$. Eikonalization of formula (\ref{2})
leads to the full shadowing correction, which reads
\beq
1 - \frac{\sigma^{\gamma^*A}_{tot}(x,Q^2)}
{A\ \sigma^{\gamma^*N}_{tot}(x,Q^2)} =
\frac{1}
{\la\sigma^{hN}_{tot}\ra}
\left\{\sqrt{\la(\sigma^{hN}_{tot})^2\ra} -
\frac{2}{A}
\int d^2b\left[1-\exp\left(-{1\over 2}
\sqrt{\la(\sigma^{hN}_{tot})^2\ra}T(b)
\right)\right]\right\}\ ,
\label{7}	
\eeq
\noi
where we assume $x$ small and fix $F_A^2=1$.
The first term in curly brackets is bigger than
the second one and both grow with $1/x$. For this reason,
the right hand side of (\ref{7}) decreases with $1/x$
more steeply than  $(1/x)^{\Delta_P(\mu^2)
- \Delta_P(Q^2)}$. Thus, addition of higher order shadowing
corrections makes vanishing of the shadowing for $x \to 0$ 
even stronger.

In order to estimate $q_L$ in (\ref{2})-(\ref{3}) we assumed
$\la m_h^2\ra \approx Q^2$. This may not be a good approximation
for so called triple-Pomeron term in shadowing, which
provides a mass distribution in diffractive dissociation
$\propto 1/m_h^2$, not steep enough to neglect the high-mass tail.
The nuclear formfactor in Gaussian form, convoluted with this mass 
distribution, results in a modified formfactor
\beq
\widetilde F_A^2(x) = 
\frac{Ei\left(-q_{max}^2R_A^2/3\right) -
Ei\left(-q_{min}^2R_A^2/3\right)}
{2\ln(q_{max}/q_{min})}\ ,
\label{8}
\eeq
\noi
where $Ei$ is the integral exponential function,
$q_{min} = m_N(x + m_{min}^2/2m_N\nu)$ and 
$q_{max} = m_N(x +m_{max}^2/2m_N\nu)$. 
The limit of integration over $m_h$ are $m_{min}$ and 
$m_{max}$. In contrast to
formfactor (\ref{3}), the modified  one (\ref{8})
grows logarithmically with $1/x$. However, with a reasonable
choice of the mass interval, this growth does not stop
the power decrease (\ref{5}) of the shadowing correction,
even if the triple Pomeron contribution (i.e. gluon fusion)
dominates nuclear shadowing.

Summarising, we predict the unusual phenomenon of vanishing nuclear
shadowing for $x\to 0$ at fixed large $Q^2$. 

\medskip

{\bf Acknowledgement:} We are grateful to Antje Br\"ull, 
who provided us with the
preliminary results of the NMC experiment for nuclear
shadowing.

}}
\setlength{\baselineskip} {5pt}

\end{document}